\documentclass[a4paper, oneside]{article}

\usepackage[utf8]{inputenc}
\usepackage[T1]{fontenc}

\usepackage[british]{babel}
\usepackage{csquotes}
\usepackage[colorlinks=true]{hyperref}
\usepackage{pdfpages,stmaryrd,amsmath,amssymb,xcolor}
\usepackage{tikz}
\usetikzlibrary{patterns}

\makeatletter
\usepgfmodule{decorations}%

\pgfdeclaredecoration{zigzag}{up from center}{%
  \state{up from center}[width=+.5\pgfdecorationsegmentlength, next state=big down]
  {
    \pgfpathlineto{\pgfqpoint{.25\pgfdecorationsegmentlength}{\pgfdecorationsegmentamplitude}}
  }%
  \state{big down}[switch if less than=+.5\pgfdecorationsegmentlength to center finish,
                   width=+.5\pgfdecorationsegmentlength,
                   next state=big up]
  {
    \pgfpathlineto{\pgfqpoint{.25\pgfdecorationsegmentlength}{-\pgfdecorationsegmentamplitude}}
  }%
  \state{big up}[switch if less than=+.5\pgfdecorationsegmentlength to center finish,
                 width=+.5\pgfdecorationsegmentlength,
                 next state=big down]
  {
    \pgfpathlineto{\pgfqpoint{.25\pgfdecorationsegmentlength}{\pgfdecorationsegmentamplitude}}
  }%
  \state{center finish}[width=0pt, next state=final]{
    \pgfpathlineto{\pgfpointorigin}
  }%
  \state{final}
  {
    \pgfpathlineto{\pgfpointdecoratedpathlast}
  }%
}%

\pgfdeclaredecoration{saw}{initial}
{%
  \state{initial}[auto end on length=+\pgfdecorationsegmentlength,
                  auto corner on length=+\pgfdecorationsegmentlength,
                  width=+\pgfdecorationsegmentlength]
  {
    \pgfpathlineto{\pgfqpoint{\pgfdecorationsegmentlength}{\pgfdecorationsegmentamplitude}}
    \pgfpathlineto{\pgfqpoint{\pgfdecorationsegmentlength}{0pt}}
  }%
  \state{final}
  {}%
}%

\pgfdeclaredecoration{random steps}{start}
{%
  \state{start}[width=+0pt,next state=step,persistent precomputation=\pgfdecoratepathhascornerstrue]{}%
  \state{step}[auto end on length=1.5\pgfdecorationsegmentlength,
               auto corner on length=1.5\pgfdecorationsegmentlength,
               width=+\pgfdecorationsegmentlength]
  {
    \pgfpathlineto{
      \pgfpointadd
      {\pgfpoint{\pgfdecorationsegmentlength}{0pt}}
      {\pgfpoint{rand*\pgfdecorationsegmentamplitude}{rand*\pgfdecorationsegmentamplitude}}
    }
  }%
  \state{final}
  {}%
}%

\pgfdeclaremetadecoration{straight zigzag}{line to}{%
  \state{line to}[width=\pgfmetadecorationsegmentlength, next state=zigzag]
  {
    \decoration{curveto}
  }%
  \state{zigzag}[width=\pgfmetadecorationsegmentlength, next state=line to]
  {
    \decoration{zigzag}
  }%
  \state{final}
  {
    \decoration{curveto}
  }%
}%

\pgfdeclaredecoration{bent}{bent}
{%
  \state{bent}[width=+\pgfdecoratedinputsegmentremainingdistance]
  {
    \pgfpathcurveto
    {\pgfqpoint{\pgfdecorationsegmentaspect\pgfdecoratedinputsegmentremainingdistance}{\pgfdecorationsegmentamplitude}}
    {\pgfpointadd{\pgfqpoint{\pgfdecoratedinputsegmentremainingdistance}{0pt}}
       {\pgfqpoint{-\pgfdecorationsegmentaspect\pgfdecoratedinputsegmentremainingdistance}{\pgfdecorationsegmentamplitude}}}
    {\pgfqpoint{\pgfdecoratedinputsegmentremainingdistance}{0pt}}
  }%
  \state{final}
  {}%
}%

\pgfdeclaredecoration{snake}{initial}
{%
  \state{initial}[switch if less than=+.625\pgfdecorationsegmentlength to final,
                  width=+.3125\pgfdecorationsegmentlength,
                  next state=down]
  {
    \pgfpathcurveto
    {\pgfqpoint{.125\pgfdecorationsegmentlength}{0pt}}
    {\pgfqpoint{.1875\pgfdecorationsegmentlength}{\pgfdecorationsegmentamplitude}}
    {\pgfqpoint{.3125\pgfdecorationsegmentlength}{\pgfdecorationsegmentamplitude}}
  }%
  \state{down}[switch if less than=+.8125\pgfdecorationsegmentlength to end down,
               width=+.5\pgfdecorationsegmentlength,
               next state=up]
  {
    \pgfpathcosine{\pgfqpoint{.25\pgfdecorationsegmentlength}{-1\pgfdecorationsegmentamplitude}}
    \pgfpathsine{\pgfqpoint{.25\pgfdecorationsegmentlength}{-1\pgfdecorationsegmentamplitude}}
  }%
  \state{up}[switch if less than=+.8125\pgfdecorationsegmentlength to end up,
             width=+.5\pgfdecorationsegmentlength,
             next state=down]
  {
    \pgfpathcosine{\pgfqpoint{.25\pgfdecorationsegmentlength}{\pgfdecorationsegmentamplitude}}
    \pgfpathsine{\pgfqpoint{.25\pgfdecorationsegmentlength}{\pgfdecorationsegmentamplitude}}
  }%
  \state{end down}[width=+.3125\pgfdecorationsegmentlength,
                   next state=final]
  {
    \pgfpathcurveto
    {\pgfqpoint{.125\pgfdecorationsegmentlength}{\pgfdecorationsegmentamplitude}}
    {\pgfqpoint{.1875\pgfdecorationsegmentlength}{0pt}}
    {\pgfqpoint{.3125\pgfdecorationsegmentlength}{0pt}}
  }%
  \state{end up}[width=+.3125\pgfdecorationsegmentlength,
                 next state=final]
  {
    \pgfpathcurveto
    {\pgfqpoint{.125\pgfdecorationsegmentlength}{-\pgfdecorationsegmentamplitude}}
    {\pgfqpoint{.1875\pgfdecorationsegmentlength}{0pt}}
    {\pgfqpoint{.3125\pgfdecorationsegmentlength}{0pt}}
  }%
  \state{final}
  {
    \pgfpathlineto{\pgfpointdecoratedpathlast}
  }%
}%

\pgfdeclaredecoration{coil}{coil}
{%
  \state{coil}[switch if less than=%
    1.5\pgfdecorationsegmentlength+%
    \pgfdecorationsegmentaspect\pgfdecorationsegmentamplitude+%
    \pgfdecorationsegmentaspect\pgfdecorationsegmentamplitude to last,
               width=+\pgfdecorationsegmentlength]
  {
    \pgfpathcurveto
    {\pgfpoint@oncoil{0    }{ 0.555}{1}}
    {\pgfpoint@oncoil{0.445}{ 1    }{2}}
    {\pgfpoint@oncoil{1    }{ 1    }{3}}
    \pgfpathcurveto
    {\pgfpoint@oncoil{1.555}{ 1    }{4}}
    {\pgfpoint@oncoil{2    }{ 0.555}{5}}
    {\pgfpoint@oncoil{2    }{ 0    }{6}}
    \pgfpathcurveto
    {\pgfpoint@oncoil{2    }{-0.555}{7}}
    {\pgfpoint@oncoil{1.555}{-1    }{8}}
    {\pgfpoint@oncoil{1    }{-1    }{9}}
    \pgfpathcurveto
    {\pgfpoint@oncoil{0.445}{-1    }{10}}
    {\pgfpoint@oncoil{0    }{-0.555}{11}}
    {\pgfpoint@oncoil{0    }{ 0    }{12}}
  }%
  \state{last}[width=.5\pgfdecorationsegmentlength+%
    \pgfdecorationsegmentaspect\pgfdecorationsegmentamplitude+%
    \pgfdecorationsegmentaspect\pgfdecorationsegmentamplitude,next state=final]
  {
    \pgfpathcurveto
    {\pgfpoint@oncoil{0    }{ 0.555}{1}}
    {\pgfpoint@oncoil{0.445}{ 1    }{2}}
    {\pgfpoint@oncoil{1    }{ 1    }{3}}
    \pgfpathcurveto
    {\pgfpoint@oncoil{1.555}{ 1    }{4}}
    {\pgfpoint@oncoil{2    }{ 0.555}{5}}
    {\pgfpoint@oncoil{2    }{ 0    }{6}}
  }%
  \state{final}
  {
    \pgfpathlineto{\pgfpointdecoratedpathlast}
  }%
}%

\def\pgfpoint@oncoil#1#2#3{%
  \pgf@x=#1\pgfdecorationsegmentamplitude%
  \pgf@x=\pgfdecorationsegmentaspect\pgf@x%
  \pgf@y=#2\pgfdecorationsegmentamplitude%
  \pgf@xa=0.083333333333\pgfdecorationsegmentlength%
  \advance\pgf@x by#3\pgf@xa%
}%

\pgfdeclaredecoration{bumps}{initial}
{%
  \state{initial}[auto end on length=+.51\pgfdecorationsegmentlength,
                  auto corner on length=+.51\pgfdecorationsegmentlength,
                  width=+.5\pgfdecorationsegmentlength]
  {
    \pgfpathcurveto
    {\pgfqpoint{0pt}{.555\pgfdecorationsegmentamplitude}}
    {\pgfqpoint{0.11125\pgfdecorationsegmentlength}{\pgfdecorationsegmentamplitude}}
    {\pgfqpoint{.25\pgfdecorationsegmentlength}{\pgfdecorationsegmentamplitude}}
    \pgfpathcurveto
    {\pgfqpoint{.38875\pgfdecorationsegmentlength}{\pgfdecorationsegmentamplitude}}
    {\pgfqpoint{.5\pgfdecorationsegmentlength}{.5\pgfdecorationsegmentamplitude}}
    {\pgfqpoint{.5\pgfdecorationsegmentlength}{0\pgfdecorationsegmentamplitude}}
  }%
  \state{final}
  {
    \pgfpathlineto{\pgfpointdecoratedpathlast}
  }%
}%

\makeatother
\usetikzlibrary{decorations.pathmorphing}

\title{Non-Robustness of the Zero-Temperature-Limit \\ Gibbs Measures to Perturbations of the Potential}
\author{Gayral Léo, Sablik Mathieu}
\date{}

\usepackage[a4paper]{geometry}
\usepackage{fullpage}

\usepackage{float}
\usepackage{caption} 

\usepackage[backend=biber,style=alphabetic,maxbibnames=9,language=british,useprefix=false]{biblatex}

\usepackage{amsthm}
\theoremstyle{plain}
\newtheorem{theorem}{Theorem}
\newtheorem{proposition}[theorem]{Proposition}

\newtheorem{definition}[theorem]{Definition}

\usepackage{tcolorbox}
\tcbset{sharp corners=all,frame empty=True}

\newcommand{\ie}{\emph{i.e.}\ }

\renewcommand{\epsilon}{\varepsilon}
\renewcommand{\phi}{\varphi}

\usepackage{amssymb}
\newcommand{\A}{\mathcal{A}}

\newcommand{\F}{\mathcal{F}}
\newcommand{\G}{\mathcal{G}}
\newcommand{\M}{\mathcal{M}}
\newcommand{\T}{\mathbf{T}}

\usepackage{dsfont}
\newcommand{\N}{\mathds{N}}
\newcommand{\Z}{\mathds{Z}}
\newcommand{\Q}{\mathds{Q}}
\newcommand{\R}{\mathds{R}}

\newcommand{\dense}{\mathfrak{D}}

\DeclareMathOperator*{\Acc}{Acc}
\newcommand{\acc}[2][]{\Acc_{#1}\left(#2\right)}

\usepackage{mathtools}
\DeclarePairedDelimiter\abs{|}{|}

\newcommand{\symb}[1]{\mathtt{#1}}

\newcommand{\robinsonone}[3]{
\begin{scope}[xshift=#1cm,yshift=#2cm,rotate=#3]
 \draw[fill=gray!25] (0.4,0.5)--(0.4,0.4)--(0.5,0.4)-- (0.5,0.1)--(0.3,0)--(0.5,-0.1) --(0.5,-0.4)--(0.4,-0.4)--(0.4,-0.5)--(0.1,-0.5)--(0,-0.7)--(-0.1,-0.5)--(-0.4,-0.5)--(-0.4,-0.4)--(-0.5,-0.4)-- (-0.5,-0.1)--(-0.3,0)--(-0.5,0.1)--(-0.5,0.4)--(-0.4,0.4)--(-0.4,0.4)--(-0.4,0.4)--(-0.4,0.5)--(-0.1,0.5)--(0,0.3)--(0.1,0.5) --cycle;
\end{scope}
}

\newcommand{\robinsontwo}[3]{
\begin{scope}[xshift=#1cm,yshift=#2cm,rotate=#3]
 \draw[fill=gray!25] (0.4,0.5)--(0.4,0.4)--(0.5,0.4)--(0.5,0.2)--(0.3,0)--(0.5,0)--(0.5,-0.4)--(0.4,-0.4)--(0.4,-0.4)--(0.4,-0.4)--(0.4,-0.5)--(0.1,-0.5)--(0,-0.7)--(-0.1,-0.5)--(-0.4,-0.5)--(-0.4,-0.4)--(-0.5,-0.4)-- (-0.5,0) -- (-0.3,0) --(-0.5,0.2) -- (-0.5,0.4)--(-0.4,0.4)--(-0.4,0.4)--(-0.4,0.4)--(-0.4,0.5) --(-0.1,0.5)--(0,0.3)--(0.1,0.5)--cycle;
  \draw[red,line width = 2pt] (0.5,0)--(-0.5,0);
\end{scope}
}

\newcommand{\robinsonleftone}[3]{
\begin{scope}[xshift=#1cm,yshift=#2cm,rotate=#3]
 \draw[fill=gray!25] (0.4,0.5)--(0.4,0.4)--(0.5,0.4)--(0.5,0.1)--(0.3,0)--(0.5,-0.1) --(0.5,-0.4)--(0.4,-0.4)--(0.4,-0.5)-- (0.2,-0.5)--(0,-0.7)--(0,-0.5) --(-0.4,-0.5)--(-0.4,-0.4)--(-0.5,-0.4)--(-0.5,-0.1)--(-0.3,0)--(-0.5,0.1)-- (-0.5,0.4)--(-0.4,0.4)--(-0.4,0.4)--(-0.4,0.4)--(-0.4,0.5) --(0,0.5)--(0,0.3)--(0.2,0.5)--cycle;
   \draw[red,line width = 2pt] (0,0.5)--(0,-0.5);
\end{scope}
}

\newcommand{\robinsonlefttwo}[3]{
\begin{scope}[xshift=#1cm,yshift=#2cm,rotate=#3]
 \draw[fill=gray!25] (0.4,0.5)--(0.4,0.4)--(0.5,0.4)--(0.5,0.2)--(0.3,0)--(0.5,0)--(0.5,-0.4)--(0.4,-0.4)--(0.4,-0.4)--(0.4,-0.4)--(0.4,-0.5)-- (0.2,-0.5)--(0,-0.7)--(0,-0.5)--(-0.4,-0.5)--(-0.4,-0.4)--(-0.5,-0.4)--(-0.5,0) -- (-0.3,0) --(-0.5,0.2)-- (-0.5,0.4) --(-0.4,0.4)--(-0.4,0.4)--(-0.4,0.4)--(-0.4,0.5)  --(0,0.5)--(0,0.3)--(0.2,0.5)--cycle;
   \draw[red,line width = 2pt] (0,0.5)--(0,-0.5);
  \draw[red,line width = 2pt] (0.5,0)--(-0.5,0);
\end{scope}
}

\newcommand{\robinsonrightone}[3]{
\begin{scope}[xshift=#1cm,yshift=#2cm,rotate=#3]
 \draw[fill=gray!24] (0.4,0.5)--(0.4,0.4)--(0.5,0.4)--(0.5,0.1)--(0.3,0)--(0.5,-0.1) --(0.5,-0.4)--(0.4,-0.4)--(0.4,-0.5)--(0,-0.5)--(0,-0.7)--(-0.2,-0.5) --(-0.4,-0.5)--(-0.4,-0.4)--(-0.5,-0.4)--(-0.5,-0.1)--(-0.3,0)--(-0.5,0.1)-- (-0.5,0.4)--(-0.4,0.4)--(-0.4,0.4)--(-0.4,0.4)--(-0.4,0.5) --(-0.2,0.5)--(0,0.3)--(0,0.5)--cycle;
   \draw[red,line width = 2pt] (0,0.5)--(0,-0.5);
\end{scope}
}

\newcommand{\robinsonrighttwo}[3]{
\begin{scope}[xshift=#1cm,yshift=#2cm,rotate=#3]
 \draw[fill=gray!25] (0.4,0.5)--(0.4,0.4)--(0.5,0.4)--(0.5,0.2)--(0.3,0)--(0.5,0)--(0.5,-0.4)--(0.4,-0.4)--(0.4,-0.4)--(0.4,-0.4)--(0.4,-0.5)--(0,-0.5)--(0,-0.7)--(-0.2,-0.5)--(-0.4,-0.5)--(-0.4,-0.4)--(-0.5,-0.4)--(-0.5,0) -- (-0.3,0) --(-0.5,0.2)-- (-0.5,0.4) --(-0.4,0.4)--(-0.4,0.4)--(-0.4,0.4)--(-0.4,0.5)   --(-0.2,0.5)--(0,0.3)--(0,0.5)--cycle;
  \draw[red,line width = 2pt] (0,0.5)--(0,-0.5);
  \draw[red,line width = 2pt] (0.5,0)--(-0.5,0);
\end{scope}
}

\newcommand{\robinsoncorner}[3]{
\begin{scope}[xshift=#1cm,yshift=#2cm,rotate=#3]
 \draw[fill=gray!25] (0.4,0.5)--(0.4,0.6)--(0.6,0.6)--(0.6,0.4)--(0.5,0.4)-- (0.5,0.2)--(0.7,0)--(0.5,0) -- (0.5,-0.4)--(0.6,-0.4)--(0.6,-0.6)--(0.4,-0.6)--(0.4,-0.5)-- (0.1,-0.5)--(0,-0.7)--(-0.1,-0.5)--(-0.4,-0.5)--(-0.4,-0.6)--(-0.6,-0.6)--(-0.6,-0.4)--(-0.5,-0.4)-- (-0.5,-0.1)--(-0.7,0)--(-0.5,0.1)--(-0.5,0.4)--(-0.6,0.4)--(-0.6,0.6)--(-0.4,0.6)--(-0.4,0.5) -- (0,0.5)--(0,0.7)--(0.2,0.5)--cycle;
 \draw[red,line width = 2pt] (0,0.5)--(0,0)--(0.5,0);
\end{scope}
}

\newcommand{\robinsoncornerbis}[3]{
\begin{scope}[xshift=#1cm,yshift=#2cm,rotate=#3]
 \draw[fill=gray!25] (0.4,0.5)--(0.4,0.4)--(0.5,0.4)--(0.5,0.2)--(0.7,0)--(0.5,0) -- (0.5,-0.4)--(0.4,-0.4)--(0.4,-0.5)--(0.1,-0.5)--(0,-0.7)--(-0.1,-0.5)--(-0.4,-0.5)--(-0.4,-0.4)--(-0.5,-0.4)-- (-0.5,-0.1)--(-0.7,0)--(-0.5,0.1)-- (-0.5,0.4)--(-0.4,0.4)--(-0.4,0.4)--(-0.4,0.4)--(-0.4,0.5) -- (0,0.5)--(0,0.7)--(0.2,0.5)--cycle;
 \draw[red,line width = 2pt] (0,0.5)--(0,0)--(0.5,0);
\end{scope}
}

\usepackage{marvosym,fancybox}

\begin{document}

\maketitle

\begin{abstract}
The robustness of properties of a statistical physics model to slight perturbations
in the exact local interactions of the model is a very relevant philosophical question,
considering real-life measurements on which we base some models can
only ever reach a finite precision.
In this article, we will discuss this topic in a formal mathematical setting,
and notably exhibit a family of models for which
the low-temperature behaviour is highly non-robust.
\end{abstract}

\tableofcontents

\section{Introduction}

\subsection{Context}

The concept of robustness to perturbations of the interactions is pretty essential
when looking at theoretical mathematical models in order to infer properties
of real-life physics phenomena.
Indeed, if the mathematical properties of the model are highly dependent on a very careful
arrangement of coefficients in the local interactions of the model,
then beyond a metaphysical interest, very little predictions may transfer to real-life experiments.

Consider for example the Newton's law of universal gravitation:
the exact value of the gravitational constant $G$ will of course affect the solutions,
but will do so continuously,
and qualitative behaviours such as the chaoticity of the three-body problem will remain.
In this sense, one could say that this model is robust to perturbations of the parameter $G$.

Here, we are more specifically interested in the study of lattice models,
with a finite alphabet $\A$ and random configurations
on the full-shift space $\Omega_\A:=\A^{\Z^d}$ called \emph{Gibbs measures},
induced by an potential $\phi:\Omega_\A\to\R$ which corresponds to the interactions,
indicating the energy contribution of the site at the origin of the lattice.
These measures correspond to the equilibrium states
of the system at inverse temperature $\beta=\frac{1}{T}$,
represented by the shift-invariant probability measures $\mu$
that maximise the pressure $h(\mu)- \beta\mu(\phi)$, where $h(\mu)$ denotes the entropy per site. 
Notably, the low-temperature behaviour of the Gibbs measures,
the set of their zero-temperature limits (the \emph{ground states} when $\beta\to\infty$),
is of particular importance in statistical physics.

Ever since the discovery of quasicrystals in the 1980s,
finding good models explaining their formation,
or even toy models for which a quasiperiodic structure emerges at low temperatures,
has been a driving force in theoretical statistical physics.
Aperiodic tilings such as those found by Robinson~\cite{Rob71}
and Penrose~\cite{Pen79} have been good candidates for these kinds of toy models,
and their combinatorial local definitions allow for a reasonable reframing
as random perturbed tilings in statistical physics where the potential simply counts
the number of disagreements with local rules for the site at the origin.
Ideally, we would like to obtain stability results for those models at low temperatures:
for some Robinson-like tilings, it is possible to obtain a kind of structural stability
according to the model studied~\cite{Mie97},
and for the Penrose tiling,
seemingly contradictory observations were made from simulations regarding
its low-temperature convergence~\cite{TanJar90,StrDre90}.
Furthermore, we would like these models to be \emph{robust},
so that they reasonably permit the emergence of quasiperiodicity without relying on
an unreachable level of theoretical purity.

In the very general setting of a \emph{continuous} interaction potential, 
ergodic optimisation results~\cite{Mor10,EntMie20} tell us that, generically, 
there exists a unique measure that minimises the average potential (\ie the unique ground state), 
with full support and zero entropy,
and the associated statistical physics models are stable at zero temperature.
Consequently, no model associated to a periodic Subshift of Finite Type (SFT),
the Kari tiling (which has positive entropy)
nor any other quasi-periodic SFT can be expected to be robust in a meaningful way.
Moreover, for a (non-generic) dense family of continuous potentials,
we have instead uncountably many minimising measures~\cite{Shi18},
which implies some kind of universal non-robustness
for continuous potentials.
A charitable understanding of this observation is that considering all continuous potentials
allows for highly ``non-physical'' models,
hence these generic behaviours that seem to go actively against
the omnipresence of ordered behaviours in Nature
(the place, not the journal, although also in the journal).

In order to add some ``physical sense'' to the models, the conventional way
is to require (exponentially) decreasing weights
for the interactions between particles on the lattice
as their distance goes to infinity.
By doing so, we in particular know that the models cannot exhibit
a freezing phase transition~\cite{ChaKucQua25},
with a low-temperature quasicrystalline phase.
A notable subset is that of \emph{finite-range} interactions,
where $\phi$ only depends on a finite window around the origin
(\ie it is a locally constant function).
Even in this more restrictive finite-range framework, as soon as we work in dimension 2 or more,
the \emph{support} of zero-temperature limits is not necessarily robust to perturbations,
as proven for the Robinson tiling~\cite{GonQuaSie21}.
Very recently, a first robust example was found
by Oguri and Shinoda~\cite{OguShi25},
with finite-range interactions robust to small Lipschitz perturbations,
so that the support of minimising measures for small perturbations is
\emph{included} in the support of
the unique non-perturbed minimising measure.

The notion of support is very restrictive since two measures very close
for a distance compatible with the weak-$*$ topology can have entirely disjoint supports
(one only needs to remember the density of measures supported by finite periodic SFTs).
It thus seems more natural to consider robustness topologically,
relatively to the set of ground states.
In other words, a potential is robust if,
for any perturbed potential in a sufficiently small neighbourhood,
the ground states are arbitrary close to the non-perturbed ones.
Relatedly, as chaoticity and stability are (mutually exclusive) yes/no discrete properties,
and the space of potentials is a continuum,
these properties cannot behave continuously in the parameter $\phi$.
Hence, we will say that a potential $\phi$
induces a \emph{robust} chaotic (resp. stable) model
if, for any potential $\psi$ in a sufficiently small neighbourhood of $\phi$,
$\psi$ induces a chaotic (resp. stable) model.

In the context of analytic models, recent results~\cite{CorRiv23} tend to indicate
that a similar notion of \emph{low-temperature sensitive dependence} is non-robust,
not just for a carefully chosen potential but in an open neighbourhood.

\subsection{Main Result}

Following a similar track of thought, in this article, we will propose a class of tilesets
that induce non-robust low-temperature limit behaviours,
by iterating upon a previous work~\cite{GaySabTaa23}.
Interestingly, in this case, the ``zero-temperature subshift''
of minimal energy configurations will be
the same regardless of the tileset/potential (\ie in the framework of ergodic optimisation, the minimising measures are the same),
the changes will only affect the positive-temperature random behaviours
and thence the accumulation set obtained by cooling trajectories.
Current evidence suggests that non-robustness is the norm rather than the exception.
However, our new result remains relevant, as it is possible to construct
a local potential that admits any target set of measures for the ground states
(under a computable assumption) in any neighbourhood of the constructed local potential.
The result can be summarised as follows:
\begin{theorem} \label{thm:main}
Let $X$ a connected $\Pi_2$-computable compact set. Then there exists a ``universal''
potential $\phi_X$ that induces the accumulation set $X$ and such that,
for any likewise connected $\Pi_2$ set $Y$, there is a potential $\psi_Y$ such that
any perturbation $\phi_X+\epsilon \psi_Y$ ($\epsilon>0$) induces $Y$ as the accumulation set.

In particular, chaotic models occur in the neighbourhood of potentials
accumulating to a singleton (\emph{i.e.} stable models), and conversely.
Notably, the potentials $\phi_X$ are not robust in the space of finite-range potentials.
\end{theorem}

Note that, unlike for the general continuous case, where non-robustness of most structures
\emph{follows} from a genericity argument, in the finite-range cases, 
current non-robustness results are constructive and focus on specific examples,
so any question about genericity or more broadly typicality of these behaviours is still open.
In particular, the \emph{existence} of some robust model
in the finite-range or exponentially-decreasing regime
is still unknown.

Still, these results point to the fact that, in the general case,
without specific arguments tailored to a specific model,
infinite precision on the parameters of the interactions are \emph{required}
to properly address whether the model has a chaotic zero-temperature cooling limit behaviour,
among other properties.
Conversely, to realistically predict the low-temperature behaviour of a given model,
we'd need some kind of robustness property that guarantees the prediction are close to reality
because we have a good empirical estimate of the parameters.

\section{Definitions and Folklore}

In this section, we will introduce the main notions necessary to understand
the key result of our previous work with Siamak Taati~\cite{GaySabTaa23},
spanning across symbolic dynamics, statistical physics and computable analysis.

\subsection{Symbolic Dynamics}

Let $\A$ be a \emph{finite} alphabet.
We denote $\Omega_\A:=\A^{\Z^d}$ the $d$-dimensional full-shift.
A \emph{subshift} of $\Omega_\A$ is a subset $X$ invariant under any shift action
$\sigma_k:\Omega_\A\to\Omega_\A$ with $k\in\Z^d$
(defined so that $\sigma_k(\omega)_l=\omega_{k-l}$) and closed in the product topology.

Let $\F$ be a set of \emph{finite} patterns $w\in A^{I(w)}$ (with $I\subset \Z^d$ finite),
and $X_\F$ the induced subshift:
\[
X_\F:=\left\{\omega\in\Omega_\A,\forall w\in\F,\forall k\in\Z^d, \sigma_k(\omega)|_{I(w)} \neq w\right\} \, .
\]
In other words, the elements of $X_\F$ are the configurations of the full-shift that contain
no translation of any \emph{forbidden pattern} $w\in\F$.
In particular, $X$ is a \emph{Subshift of Finite Type} (SFT) if there exists
a \emph{finite} set of forbidden patterns $\F$ such that $X=X_\F$.

Let $\M\left(\Omega_\A\right)$ be the set of probability measures on the full-shift.
We endow this space with the weak-$*$ topology, \emph{i.e.} the weakest topology such that 
$\mu\in\M\left(\Omega_\A\right)\mapsto\mu([w])$ is continuous for any finite pattern $w\in\A^I$
and $[w]:=\left\{\omega,\omega|_I=w\right\}$ the corresponding cylinder set. 
We then denote $\M_\sigma(X)$ the (closed) set of shift-invariant measures
(\emph{i.e.} such that $\mu\circ \sigma_k=\mu$ for any $k\in\Z^d$) supported by the subshift $X$.

\subsection{Statistical Physics}

Now, in order to fit a statistical physics framework,
we need a configuration space and a notion of ``energy''.
This will allow us to describe a random state of the system
that realises a compromise between lowering the energy (order and structure)
and increasing the entropy (disorder and complexity),
in order to conclude on which behaviours are typical.

We are here interested in \emph{lattice models},
so the configuration space will simply be the full-shift $\Omega_\A$.
Intuitively, the total energy of a lattice will be infinite in many cases,
so we need to adapt the framework to account for this.
One way to do so is to restrict ourselves to the shift-invariant case,
where there is an explicit link between the global behaviour and the local one.
More precisely, we define the \emph{potential} function $\phi:\Omega_\A\to\R^+$,
that represents the contribution of the origin $0\in\Z^d$ of the lattice to the overall energy
(informally, the total energy is $\sum_k \phi\circ \sigma_k$).

Likewise, we can define the averaged entropy per site of a measure
$h(\mu):= \lim_{n\to\infty} \frac{1}{n^d}H\left(\mu|_{\llbracket 0,n-1\rrbracket^d}\right)$
(the limit is always well-defined in the shift-invariant case,
and doesn't really depend on the choice of increasing family
of sets that asymptotically covers the lattice), and consider the set of \emph{Gibbs measures}
$\G(\beta)$ that maximise the \emph{pressure} function
$p_\beta:\mu\mapsto h(\mu)-\beta \times \int \phi \mathrm{d}\mu$
at inverse temperature $\beta \in\R^+$.

One of the core questions in statistical physics is that
of the low-temperature limit behaviour of models,
to explain the formation of quasicrystals and other complex materials.
Thus, we define the accumulation set of \emph{ground states}
$\G(\infty):=\text{Acc}_{\beta\to\infty} \G(\beta) = \left\{
\lim_{n\to\infty}\mu_{\beta_n},\beta_n\to\infty,\mu_{\beta_n}\in\G\left(\beta_n\right)\right\}$,
the set of \emph{all} the possible zero-temperature limit measures.

We say that a model is \emph{stable} when all the trajectories
$\left(\mu_\beta\right)_{\beta\in\R^+}$ converge as $\beta\to\infty$,
or equivalently when $\G(\infty)$ is a singleton.
Conversely, we say that a model is (strongly) \emph{chaotic}
when all the trajectories \emph{don't} converge.

\subsection{Computable Analysis}

The last important notion we must overview is computability theory,
with a certain focus on computable analysis,
\ie how to adapt computability from integers (or more generally countable sets)
to general topological or metric (uncountable) spaces.

First of all, regarding computability on integers,
most automatic or intuitive frameworks (such as pseudo-code algorithms)
have the same ``computing power'',
and the one we use is that of \emph{Turing machines}, embedded within tilings.
Formally, a Turing machine $M$ is a tuple $\left(Q,q_0,Q_F,\A,\Gamma,\delta\right)$.
$Q$ is the finite set of internal states of the machine,
$q_0\in Q$ is its initial state, and $Q_F\subset Q$ the set of \emph{final} states.
$\A$ is the input alphabet of the machine, and $\Gamma\supsetneq \A$ the \emph{tape} alphabet.
Lastly, $\delta:Q\times\Gamma\to Q\times\Gamma\times\{\pm 1\}$
is the \emph{transition} function.
The machine $M$ receives a finite word $w_0\in\A^*$
as its input (written on the semi-infinite tape as $w_0\#^\infty\in\Gamma^\N$),
and then follows through a trajectory
according to its transition function $\delta$,
until it possibly reaches a halting state in $Q_F$.
More precisely, initially $M$ is in a configuration
$\left(q_0,w_0,0\right)$,
and we have the transitions
$\left(q_t,w_t,n_t\right)\mapsto\left(q_{t+1},w_{t+1},n_{t+1}\right)$
defined inductively by
$\left(q_{t+1},b,d\right):=\delta\left(q_t,w_t\left[n_t\right]\right)$,
$w_{t+1}$ is equal to $w_t$ except in position $n_t$ where
$w_{t+1}\left[n_t\right]=b$,
and $n_{t+1}=n_t+d$ (with the convention that $0-1=0$ in $\N$).
If the machine reaches a halting state $q_t\in Q_F$,
then we simply stall the trajectory,
\ie $\left(q_{t+1},w_{t+1},n_{t+1}\right):=\left(q_t,w_t,n_t\right)$.
By choosing an encoding of integers in base $|\A|$,
we have a correspondence between integers in $\N$ and words in $\A^*$,
that allows us to see a Turing machine as a partial function on $\N$ in certain cases
(with $n$ in the support of the function \emph{iff} the machine halts).
Likewise, for any countable spaces $X$ and $Y$ such as $\Q$,
for which we can have an intuitive/explicit correspondence with $\A^*$ or $\N$,
we can have a similar notion of computable functions from $X$ to $Y$.

Computability as defined above is intrinsically related to countable objects.
However, analysis is usually performed on uncountable spaces such as $\R$,
that don't fit this framework: that's where \emph{computable analysis} comes into play.
A \emph{computable metric space} is a quadruplet 
$\left(X,d,\dense=\left(z_n\right)_{n\in\N},\delta\right)$
where $(X,d)$ is a metric space, $\dense$ is a dense family,
and $\delta:\dense^2\times\N\to\Q$ is a computable map such that,
for any $x,y\in\dense$ and $n\in\N$,
we have $\left|\delta(x,y,n)-d(x,y)\right|\leq \frac{1}{2^n}$.
Put simply, it means that the metric $d$ is computable in some sense.
Notably, in the case of $\R$, by using $\dense=\Q$, we can directly compute $d:\Q^2\to\Q$.

If $(X,d,\dense,\delta)$ is a computable space,
then the set of dyadic probability measures on $\dense$
is a dense family in $\M(X)$ the space of probability measures on $X$ (with an arbitrary convenient choice of distance that induces the weak-$*$ topology).

The number of quantifiers necessary until we reach an ``easy'' problem
(easy is here to be understood as \emph{computable} in the usual way on countable spaces)
becomes a way to quantify the complexity of an undecidable problem
to make it fit in the \emph{arithmetical hierarchy}.
We will not detail here the intricacies of this hierarchy
of $\left(\Sigma_k,\Pi_k\right)_{k\in\N}$ sets,
but we will underline another characterisation of $\Pi_2$-computable sets specifically.
For the rest of this article, we will define a $\Pi_2$ set
$X$ as the accumulation set of a computable sequence,
\emph{i.e.} $X=\acc{x_n}$
with $x:n\in\N\mapsto x_n\in \dense$ computable.

Working within these frameworks, we proved the following result:

\begin{theorem}[\cite{GaySabTaa23}] \label{thm:old}
There exists a two-dimensional subshift of finite type $\T\subset \Omega_\A$,
and a computable affine bijection $\gamma^*:\M_{\sigma}(\T)\to\M(\{\downarrow,\uparrow\}^\N)$,
such that for any $\Pi_2$-computable connected compact set 
$K\subset\M(\{\downarrow,\uparrow\}^\N)$,
there is a potential $\phi_K$ on a larger alphabet $\A_K$ such that
$\G(\infty)\subset \M_\sigma(\T)$ 
and $\gamma^*(\G(\infty))=K$. 
\end{theorem}

\section{Digest of the Construction for Theorem~\ref{thm:old}} \label{sec:old-paper}

The purpose of this section is to highlight a few key steps
and results in the construction used for Theorem~\ref{thm:old},
both in order to digest the corresponding article in a more accessible way
and to give the key ideas necessary for the variant used
in the next section to prove our new non-robustness result.
We will \emph{not} go too deep in the technicalities,
as these were already very detailed in the related article.

To prove this theorem, we use the characterisation of
a $\Pi_2$-computable set $K$ as the accumulation set $\acc{x_n}$
of a computable sequence $n\in\N\mapsto x_n\in\dense$.
More precisely, we will consider a more precise characterisation of
\emph{connected} $\Pi_2$ sets as accumulation sets for a computable sequence
$\left(x_n\right)$ such that 
$d\left(x_n,x_{n+1}\right)\underset{n\to\infty}{\longrightarrow} 0$
(see \cite[Proposition 5]{GaySabTaa23} for the details).

Thence, our goal is to build an SFT $X$,
and a computable map $\gamma:X\to \{\downarrow,\uparrow\}^\N$
that induces a (computable) affine bijection between
$\M(\{\downarrow,\uparrow\}^\N)$ and the invariant measures $\M_\sigma(X)$
(so that we can identify their $\Pi_2$-computable sets).
Moreover, for any $\Pi_2$-computable set 
$K=\acc{x_n}\subset\M(\{\downarrow,\uparrow\}^\N)$,
we want to have a set of forbidden patterns $\F$
(for which $X_\F=X$) such that the Gibbs measures
associated to the corresponding potential $\phi$ satisfy:
\[
\forall k\in\N,\forall \beta\in\left[\beta_k,\beta'_{k}\right],
d^*\left(\G(\beta),\gamma^{-1}\left(x_k\right)\right)\leq \delta_k \, ,
\]
with $\delta_k\to 0$,
$\beta_k\to\infty$ and $\beta_k'\to\infty$ in an alternating way,
such that $\beta_k\leq\beta_{k+1}\leq\beta'_k\leq\beta'_{k+1}$ for
any $k\in\N$ (\ie the intervals $\left[\beta_k,\beta'_{k}\right]$ overlap),
in which case $\G(\infty)=\gamma^{-1}(K)$.

\subsection{Entropy and Distribution of Markers in Gibbs Measures}

The first ingredient of the proof is an ergodic theorem
which quantifies the distribution of a family of finite patterns,
called \emph{ground markers}.

For $n\in\N$, denote $I_n=\llbracket 0,n-1\rrbracket^2$
the $n$-square window and $G_{I_n}$
the set of patterns on $I_n$ where the local rules are respected.

\begin{definition}[Ground Markers] \label{def:marker}
Let $\ell\in\N$. A \emph{marker set} of \emph{margin factor} $\tau$ is a set $Q\subseteq G_{I_\ell}$
of ground patterns on $I_\ell$ that satisfies the following conditions:
\begin{itemize}
    \item \emph{Non-overlapping}: For every $u,v\in Q$ and $k\in\Z^d$,
    if $u_{I_\ell\cap(-k+I_\ell)}=(\sigma^k (v))_{I_\ell\cap(-k+I_\ell)}$,
    then either $I_\ell\cap(-k+I_\ell)=\emptyset$ or $k=0$.
    \item \emph{Covering}: For every $w\in G_{I_m}$,
    with $m=(2+\tau)\ell-1$,
    there is a translation $k$ such that $(\sigma^k (w))_{I_\ell}\in Q$.
\end{itemize}
\end{definition}

Informally, a marker set is a set of square patterns
that cannot partially overlap,
and such that any admissible pattern roughly twice their size must contain
a marker somewhere inside.

\begin{theorem}[Equidistribution]~\cite[Theorem 26]{GaySabTaa23} \label{thm:equidistribution}
Let $\Phi$ be a non-negative finite-range  interaction (of range $r$) which admits null-energy configurations. There exists $C,C'>0$ such that for any $\ell$-marker set with margin factor $\tau$ denoted $Q$ (recall that $m=(2+\tau)\ell-1$)  and $\kappa,\epsilon>0$ and $n\geq 2r$, under the assumption that the following two criteria are satisfied:
\begin{itemize}
    \item Entropy: $\frac{\log_2\left(\abs{G_{I_n}}\right)}
    {\abs{I_n}}\geq (1-\kappa)\frac{\log_2(\abs{Q})}{\abs{I_\ell}}$,
    \item Temperature:
    $\frac{C}{\epsilon}\abs{I_m}\leq\beta\leq
   C'\epsilon\frac{
    \abs{I_n}}{\abs{I_n}-\abs{I_{n-2r}}}$.
\end{itemize}
Then, for every $\mu\in\G(\beta)$,
we have the following properties:
\begin{enumerate}
\item Covering: $\mu(0\triangleleft Q)\geq 1-\epsilon'$ where 
$\epsilon':=1-\frac{1-\epsilon}{(1+\tau)^2}$ and ``$0\triangleleft Q$'' 
denote the event where the origin is covered by an element of $Q$
(\ie $\{x\in \Omega_\A,\exists i\in\Z^d,0\in I+i,x_{I+i}\in Q\}$).
\item Uniformity:
$\frac{H\left(\mu_Q\right)}{\abs{I_\ell}}\geq (1-2\kappa)\frac{\log_2(\abs{Q})}{\abs{I_\ell}}-o(\epsilon,\epsilon',\kappa)$ .
\end{enumerate}
\end{theorem}

The first property simply states that markers occupy a high-density
area within a typical random configuration, that there is a high probability
of landing inside a marker by looking at a given position.
The second property means that the induced distribution on markers
(conditionally to the previous high-probability event of landing inside one)
for any Gibbs measure in the temperature range has a high entropy,
thus it must be close to the uniform distribution on $Q$.
Once put together and digested, these properties imply that any measure
$\mu\in \G(\beta)$ (for $\beta$ in a certain interval)
is in a neighbourhood of $\overline{\lambda_Q}$ in the weak-$*$ topology,
with $\overline{\lambda_Q}$ the shift-invariant measure induced by a grid
of independently random markers uniformly distributed in $Q$.
From there, the strategy is to build a tileset that induces a \emph{sequence}
of sets $Q_k$
of increasingly large markers,
that must satisfy the following properties:
\begin{enumerate}
\item the margin factor $\tau_k$ must go to $0$ (as well as $\epsilon_k$ and $\kappa_k$) in such a way that $\epsilon'_k$ goes to $0$,
which notably allows us to asymptotically neglect what happens outside of markers;
\item the normalised entropies of the marker sets $Q_k$ go to $0$ slowly enough,
so that the first assumption of Theorem~\ref{thm:equidistribution}
can be satisfied;
\item the size of the markers increases fast-enough, so that a bigger marker from $Q_{k+1}$ still mostly induces the same behaviour as one in $Q_k$, 
which allows the inverse temperature intervals
$\left[\beta_k,\beta'_{k}\right]$ to overlap;
\item the distribution $\lambda_{Q_k}$,
through the structure of the markers in $Q_k$,
must be controllable by local rules,
and Turing machines are an appropriate
tool for the task.
\end{enumerate}

\subsection{Construction of a Simulating Tileset Inducing Markers}
\label{sec:simulating-tileset}

\subsubsection{Multiscale Markers in the Robinson Tiling}

First, we need a structural baseline inducing a family of
increasingly large marker sets, in which we will implement computations.
The Robinson tiling~\cite{Rob71} is a good candidate,
as it induces self-similar macro-tiles at every scale,
and permits a well known embedding of Turing machines
(historically used to provide a simple proof of undecidability for the domino problem).
The tileset is given by the following $9$ tiles and their right angle rotations:
\begin{figure}[!ht]
\centering
 \begin{tikzpicture}[scale=0.5]
\robinsonone{0}{0}{0} \robinsonleftone{2}{0}{0} \robinsonrightone{4}{0}{0} \robinsoncorner{6}{0}{0} \robinsontwo{8}{0}{0} \robinsonlefttwo{10}{0}{0} \robinsonrighttwo{12}{0}{0}\robinsoncornerbis{14}{0}{0} 
\end{tikzpicture}
\caption{The Robinson tileset} \label{fig:rob-tiles}
\end{figure}
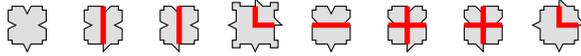

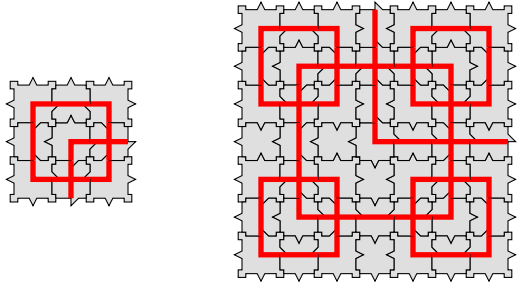
\begin{figure}[!ht]
\centering
\begin{tikzpicture}[scale=0.5]
\robinsoncorner{-1}{-1}{0}  \robinsoncorner{1}{-1}{90}  \robinsoncorner{1}{1}{180}  \robinsoncorner{-1}{1}{-90}
 \robinsoncornerbis{0}{0}{-90} 
\robinsonlefttwo{0}{-1}{0} \robinsonrighttwo{1}{0}{90} \robinsontwo{0}{1}{180} \robinsontwo{-1}{0}{-90}

\begin{scope}[xshift=8cm]
\begin{scope}[xshift=-2cm,yshift=-2cm,rotate=90]
\robinsoncorner{-1}{-1}{0}  \robinsoncorner{1}{-1}{90}  \robinsoncorner{1}{1}{180}  \robinsoncorner{-1}{1}{-90}
\robinsoncornerbis{0}{0}{-90} 
\robinsonlefttwo{0}{-1}{0} \robinsonrighttwo{1}{0}{90} \robinsontwo{0}{1}{180} \robinsontwo{-1}{0}{-90}
 \end{scope}

 \begin{scope}[xshift=2cm,yshift=-2cm,rotate=180]
\robinsoncorner{-1}{-1}{0}  \robinsoncorner{1}{-1}{90}  \robinsoncorner{1}{1}{180}  \robinsoncorner{-1}{1}{-90}
\robinsoncornerbis{0}{0}{-90} 
\robinsonlefttwo{0}{-1}{0} \robinsonrighttwo{1}{0}{90} \robinsontwo{0}{1}{180} \robinsontwo{-1}{0}{-90} 
\end{scope}

 \begin{scope}[xshift=2cm,yshift=2cm,rotate=-90]
\robinsoncorner{-1}{-1}{0}  \robinsoncorner{1}{-1}{90}  \robinsoncorner{1}{1}{180}  \robinsoncorner{-1}{1}{-90}
\robinsoncornerbis{0}{0}{-90} 
\robinsonlefttwo{0}{-1}{0} \robinsonrighttwo{1}{0}{90} \robinsontwo{0}{1}{180} \robinsontwo{-1}{0}{-90}
 \end{scope}
 
 \begin{scope}[xshift=-2cm,yshift=2cm,rotate=0]
\robinsoncorner{-1}{-1}{0}  \robinsoncorner{1}{-1}{90}  \robinsoncorner{1}{1}{180}  \robinsoncorner{-1}{1}{-90}
\robinsoncornerbis{0}{0}{-90} 
\robinsonlefttwo{0}{-1}{0} \robinsonrighttwo{1}{0}{90} \robinsontwo{0}{1}{180} \robinsontwo{-1}{0}{-90}
 \end{scope}
 
 \robinsoncornerbis{0}{0}{-90}  \robinsoncornerbis{0}{0}{180} 
 \robinsoncornerbis{0}{0}{90} 
 \robinsoncornerbis{0}{0}{0} 
\robinsonleftone{1}{0}{90}\robinsonlefttwo{2}{0}{90}\robinsonleftone{3}{0}{90}
\robinsonrightone{0}{1}{180}\robinsonrighttwo{0}{2}{180}\robinsonrightone{0}{3}{180}
\robinsonone{-1}{0}{-90}\robinsontwo{-2}{0}{-90}\robinsonone{-3}{0}{-90}
\robinsonone{0}{-1}{0}\robinsontwo{0}{-2}{0}\robinsonone{0}{-3}{0}
\end{scope}
\end{tikzpicture}
\caption{A 2-macro-tile and a 3-macro-tile}
\label{fig.SuperTileRobi}
\end{figure}

The self-similar hierarchical structure of the Robinson tiling works as follows.
In Figure~\ref{fig:rob-tiles}, the rotations of the (fourth from the left) tile with bumpy corners initialise the induction as $1$-macro-tiles. The lines on top of them can be assembled to draw a square, and filling the central cross draws a bigger red corner,
a $2$-macro-tile, as seen in Figure~\ref{fig.SuperTileRobi}.
Likewise, we can draw a square with $2$-macro-tiles (resp. $n$) and then
fill the cross, which draws a bigger corner, a $3$-macro-tiles (resp. $n+1$).

The Robinson SFT $\T_{\mathcal{R}obi}$ (\ie the set of admissible tilings of $\Z^2$)
is non-empty, aperiodic, and uniquely ergodic. For any scale $n$, the (four) $n$-macro-tiles are non-overlapping squares of length $l_n=2^n-1$.
By adding some more lines on the tiles to strengthen the local rules,
we obtain a variant of this SFT such that any admissible square pattern
of length $2\ell_n+5$ contains an $n$-macro-tile (\ie the margin factor is
$\tau=\frac{6}{\ell_n}$).

Formally, from now on, $P_k$ will denote the set of Robinson $k$-macro-tiles
(that are locally admissible on the other yet-to-be-discussed layers).
In particular, $\left(P_k\right)_{k\in\N}$
forms a family of increasingly large marker sets.
Thereon, we need to add yet more structure to the markers,
so that a marker set produces a certain distribution on words encoded in the markers.
For a given temperature, the local rules force the apparition of some scale of markers,
and so a given distribution on words.

\subsubsection{Alternating Structure for Turing Computations}

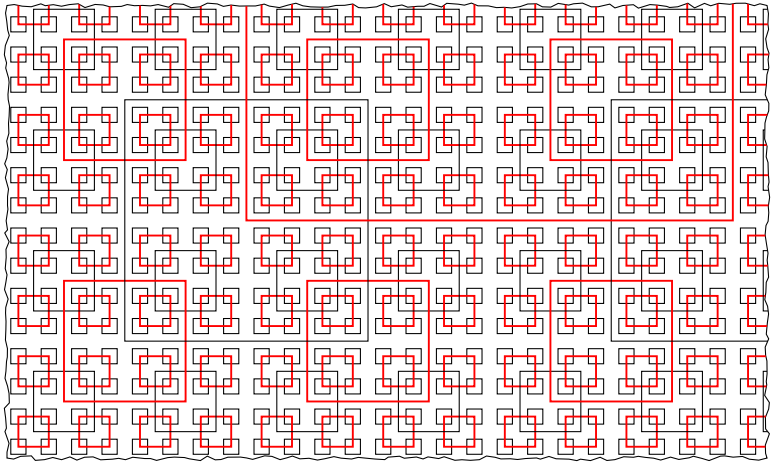
\begin{figure}[!ht]
\centering
\begin{tikzpicture}[scale=0.10]
\clip[draw,decoration={random steps, segment length=3pt, amplitude=1pt}]decorate{(-0.5,-0.5) rectangle++(100,60)}--cycle;
\foreach \x in {0,4,...,100}{\foreach \y in {0,4,...,60}{
    \draw[black] (\x,\y) rectangle ++(2,2);
}}
\foreach \x in {1,9,...,100}{\foreach \y in {1,9,...,60}{
    \draw[line width=0.25mm, red] (\x,\y) rectangle ++(4,4);
}}
\foreach \x in {3,19,...,100}{\foreach \y in {3,19,...,60}{
    \draw[black] (\x,\y) rectangle ++(8,8);
}} 
\foreach \x in {7,39,...,100}{\foreach \y in {7,39,...,60}{
    \draw[line width=0.25mm, red] (\x,\y) rectangle ++(16,16);
}} 
\foreach \x in {15,79,...,100}{\foreach \y in {15,79,...,60}{
    \draw[black] (\x,\y) rectangle ++(32,32);
}}
\foreach \x in {31,159,...,100}{\foreach \y in {31,159,...,30}{
    \draw[line width=0.25mm, red] (\x,\y) rectangle ++(64,64);
}}
\end{tikzpicture}
\caption{Alternating structure in Robinson macro-tiles}
\label{fig:robinson-alternating}
\end{figure}

In order to control the distribution on words with local rules,
a natural approach is to embed Turing machines in the Robinson tiling.
We use the standard embedding of a Turing machine,
as described in Robinson's own article~\cite{Rob71}.
This classical procedure distinguishes macro-tiles of even and odd scales
in $\T_{\mathcal{R}obi}$,
by using two colours on the lines and forcing them to alternate
as the lines of two consecutive scales of squares cross,
as seen on Figure~\ref{fig:robinson-alternating}.
Then, the Turing machines are encoded on a sparse area
in each Red-coloured square (while avoiding smaller Red squares).

\subsubsection{Hot and Frozen Areas}

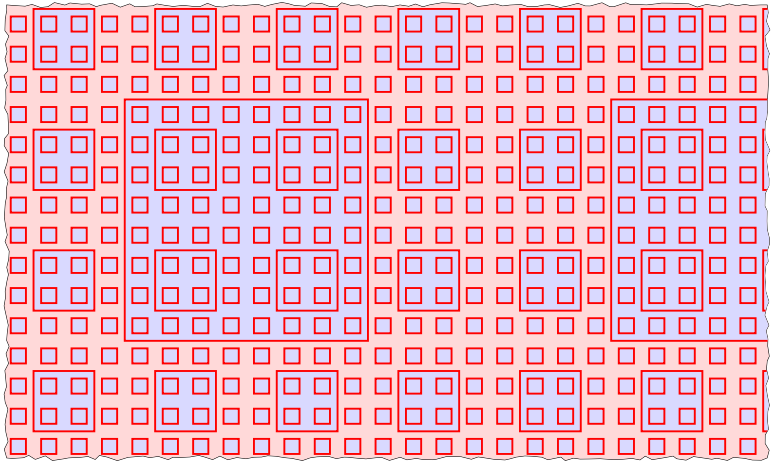
\begin{figure}[!ht]
\centering
\begin{tikzpicture}[scale=0.1]
\clip[draw,decoration={random steps, segment length=3pt, amplitude=1pt}]decorate{(-0.5,-0.5) rectangle++(100,60)}--cycle;
\fill[red!15] (-1,-1)rectangle++(101,61);
\foreach \x in {15,79,...,100}{\foreach \y in {15,79,...,60}{
    \fill[blue!15] (\x,\y) rectangle ++(32,32);
    \draw[line width=0.25mm, red] (\x,\y) rectangle ++(32,32);
}}
\foreach \x in {3,19,...,100}{\foreach \y in {3,19,...,60}{
    \fill[blue!15] (\x,\y) rectangle ++(8,8);
    \draw[line width=0.25mm, red] (\x,\y) rectangle ++(8,8);
}}
\foreach \x in {0,4,...,100}{\foreach \y in {0,4,...,60}{
    \fill[blue!15] (\x,\y) rectangle ++(2,2);
    \draw[line width=0.25mm, red] (\x,\y) rectangle ++(2,2);
}}
\end{tikzpicture}
\caption{Hot area with Blocking squares of various scales inside}
\label{fig:Blocking}
\end{figure}

We now add a three-phased layer $\A_{\mathrm{phase}}=\{\symb{F},\symb{B},\symb{H}\}$ to obtain an SFT $\T_{\F_2}\subset\T_{\mathcal{R}obi} \times\{\symb{F},\symb{B},\symb{H}\}$
with Hot and Frozen areas.
Hot areas will serve to ``produce entropy''
while Frozen areas will simulate and encode the target measures on words.
The local rules force the symbol $\symb{B}$ to only appear on Red lines,
as an interface between an outer Hot area labeled $\symb{H}$
and an inner Frozen area labeled $\symb{F}$.
As such, we call these Red squares Blocking.
Figure~\ref{fig:Blocking} represent an example of configuration,
with Hot areas in red, Frozen areas in blue,
and the symbols $\symb{B}$ represented by tiles of type
$\left\{\raisebox{-0.7mm}{\begin{tikzpicture}[scale=0.3]\draw (0,0)rectangle (1,1); \fill[blue!15] (0,0)rectangle++(1,0.5);\fill[red!15] (0,0.5)rectangle++(1,0.5);\draw[red] (0,0.5)--++(1,0);\end{tikzpicture}} \, ,
\raisebox{-0.7mm}{\begin{tikzpicture}[scale=0.3]\draw (0,0)rectangle (1,1); \fill[blue!15] (0,0)rectangle++(0.5,0.5);\fill[red!15](0,0.5)--++(0.5,0)--++(0,-0.5)--++(0.5,0)--++(0,1)--++(-1,0)--cycle;\draw[red] (0,0.5)--++(0.5,0)--++(0,-0.5);\end{tikzpicture}} \right\}$.

We remark that in a macro-tile (or a configuration of $\T_{\F_2}$), we have at most one Hot connected outer area which can contain several
Blocking squares, with all their inside being Frozen.
Alternatively, we may have an entirely Frozen macro-tile.

\subsubsection{Limiting the Scales of Blocking Squares} \label{sec:blockable}

In order for the entropy to decrease to $0$ slowly enough,
we want to control and limit which scales of macro-tiles are allowed 
o have a Blocking square (marked by the $\symb{B}$),
which we will call Blockable scales.
This process is done in two steps,
within each Red square (where it is possible to implement computations):
\begin{itemize}
\item Computation space-time diagrams at the $n$-th scale of Red squares have 
a tape of size $2^n$, from which we can \emph{procedurally} compute $n$;
\item then, we simply check if $n$ is a power of $3$,
in which case we compute the value $k$ such that $n=3^k$.
\end{itemize}

Thus, we obtain an SFT
$\T_{\F_3}\subset \T_{\F_2}\times\A_{\mathrm{scale}}^{\Z^2}$,
such that the symbol $\symb{B}$ can only appear
if the scale of the Red square is a power of $3$.

From now on, we are going to consider the marker sets
$Q_k=P_{n_k}$ with $n_k=2\times 3^k+1$,
so that $Q_k$ corresponds to the $k$-th scale
of complete Blockable Red squares.
These marker sets can be decomposed as
$Q_k=Q_k^\symb{H}\sqcup Q_k^\symb{B}\sqcup Q_k^\symb{F}$,
depending on whether the central Red square is Blocking, Hot or all-Frozen.

\subsubsection{Forcing the Density of Blocking Squares}

In order to control the density of Blocking squares,
we implement an odometer at every Blockable scale so that,
at the scale of $k$-markers, periodically, only one in $t_k\simeq\log(k)$ marker
is Blocking, and otherwise Hot.
Of course, this only applies in Hot areas, and doesn't affect Frozen markers.
We thus obtain an SFT $\T_{\F_4}\subset\T_{\F_3}\times\A_{\mathrm{Density}}^{\Z^2}$.

These Blocking squares gradually chip away the Hot area from the top down,
starting from the bigger scales, and then the smaller ones in the leftover Hot areas.
If we denote $\textrm{freq}_\symb{F}^k$ the density of Frozen areas in a markers of $Q^\symb{H}_k$,
by decomposing a marker of $Q^\symb{H}_{k+1}$
as a grid of markers from $Q^\symb{B}_k$ and $Q^\symb{H}_k$, we have the induction:
\[
\textrm{freq}_\symb{F}^{k+1}=\underset{\text{markers from }Q^\symb{B}_k}
{\underbrace{\frac{1}{t_k}\left(\frac{1}{4}+\frac{3}{4}\textrm{freq}_\symb{F}^{k}\right)}}
+\underset{\text{markers from }Q^\symb{H}_k}
{\underbrace{\frac{t_k-1}{t_k}\textrm{freq}_\symb{F}^{k}}}.
\]
We can thus rewrite $1-\textrm{freq}_\symb{F}^{k+1}=\left(1-\frac{1}{4t_{k+1}}\right)\left(1-\textrm{freq}_\symb{F}^{k}\right)$. As $t_k=O(k)$, the left factor is the general term of a null infinite product,
so $\textrm{freq}_\symb{F}^k \to 1$ and:
\begin{proposition}[{\cite[Proposition 34]{GaySabTaa23}}] \label{prop:frozen}
The SFT $T_{\F_4}$ is uniquely ergodic,
and for the invariant measure on this SFT, almost surely, all the cells of the configuration have the symbol $\symb{F}$ on the second layer.
\end{proposition}

\subsubsection{Additional Bits on Lines to Encode Words}

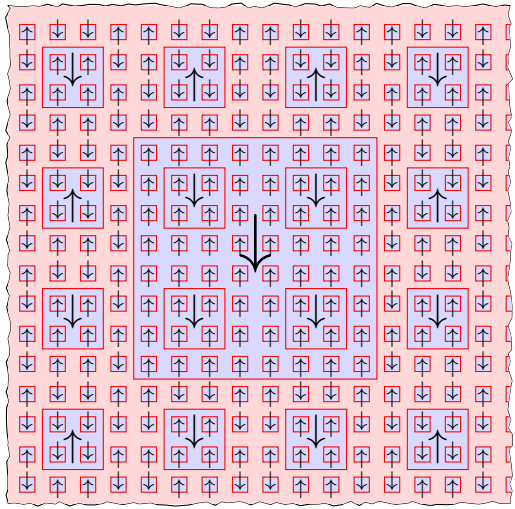
\begin{figure}[!ht]
\centering
\begin{tikzpicture}[scale=0.1]
\clip[draw,decoration={random steps, segment length=3pt, amplitude=1pt}]decorate{(-1.5,-1.5) rectangle++(66,66)}--cycle;
\fill[red!15] (-1.5,-1.5)rectangle++(101,66);
\foreach \x in {15,79,...,100}{\foreach \y in {15,79,...,60}{
    \fill[blue!15] (\x,\y) rectangle ++(32,32);
    \draw[red] (\x,\y) rectangle ++(32,32);
}}
\foreach \x in {3,19,...,100}{\foreach \y in {3,19,...,60}{
    \fill[blue!15] (\x,\y) rectangle ++(8,8);
    \draw[red] (\x,\y) rectangle ++(8,8);
}}
\foreach \x in {0,4,...,100}{\foreach \y in {0,4,...,60}{
    \fill[blue!15] (\x,\y) rectangle ++(2,2);
    \draw[red] (\x,\y) rectangle ++(2,2);
}}
\begin{footnotesize}
\foreach \x in {16,20,...,44}{\foreach \y in {16,20,...,44}{
    \draw (\x+1,\y+.5) node{$\uparrow$};
}}
\foreach \x in {0,12,48,60}{\foreach \y in {0,8,...,60}{
    \draw (\x+1,\y+1.5) node{$\downarrow$};
}}
\foreach \x in {0,12,48,60}{\foreach \y in {4,12,...,60}{
    \draw (\x+1,\y+.5) node{$\uparrow$};
}}
\foreach \x in {16,28,32,44}{\foreach \y in {0,8,48,52}{
    \draw (\x+1,\y+1.5) node{$\downarrow$};
}}
\foreach \x in {16,28,32,44}{\foreach \y in {4,12,56,60}{
    \draw (\x+1,\y+.5) node{$\uparrow$};
}}
\foreach \x in {4,8,20,24,36,40,52,56}{\foreach \y in {0,48}{
    \draw (\x+1,\y+.5) node{$\uparrow$};
}}
\foreach \x in {4,8,20,24,36,40,52,56}{\foreach \y in {12,60}{
    \draw (\x+1,\y+1.5) node{$\downarrow$};
}}
\foreach \x in {4,8,52,56}{\foreach \y in {16,32}{
    \draw (\x+1,\y+.5) node{$\uparrow$};
}}
\foreach \x in {4,8,52,56}{\foreach \y in {28,44}{
    \draw (\x+1,\y+1.5) node{$\downarrow$};
}}
\foreach \i/ \j in{0/0,48/0,48/32,32/48,16/48,0/32}{
    \begin{scope}[xshift=\i cm,yshift=\j cm]
    \foreach \x in {4,8}{\foreach \y in {4,8}{
        \draw (\x+1,\y+1.5) node{${\downarrow}$};}}
    \end{scope}
}
\foreach \i/\j in{16/0,32/0,48/16,48/48,0/48,0/16}{
    \begin{scope}[xshift=\i cm,yshift=\j cm]
    \foreach \x in {4,8}{\foreach \y in {4,8}{
        \draw (\x+1,\y+.5) node{${\uparrow}$};}}
    \end{scope}
}
\end{footnotesize}
\begin{Large}
\foreach \x in {19,35}{\foreach \y in {19,35}{
    \draw (\x+4,\y+5) node{$\downarrow$};
}}
\foreach \i/ \j in{0/0,48/0,48/32,32/48,16/48,0/32}{
    \begin{scope}[xshift=\i cm,yshift=\j cm]
        \draw (7,6) node{${\uparrow}$};
    \end{scope}
}
\foreach \i/\j in{16/0,32/0,48/16,48/48,0/48,0/16}{
    \begin{scope}[xshift=\i cm,yshift=\j cm]
        \draw (7,8) node{${\downarrow}$};
    \end{scope}
}
\end{Large}
\begin{Huge}
\draw (31,33) node{$\downarrow$};
\end{Huge}
\end{tikzpicture}
\caption{Inside each Frozen area, squares of the same scale encode the same bit}
\label{fig:words}
\end{figure}

On this new layer, we want each Blocking square to encode a binary word,
so that we can use them in computations later.
To do so, each Red line will encode an additional bit,
and these bits will locally synchronise between neighbouring squares within each Frozen area,
as illustrated in Figure~\ref{fig:words}.
This way, we will have many ``independent'' short words,
one for each Blocking square in a big marker in $Q_k^\symb{H}$.

More formally, the new SFT is $\T_{\F_5}\subset\T_{\F_4}\times(\{\uparrow_H,\downarrow_H,\bullet_H\}\times\{\uparrow_V,\downarrow_V,\bullet_V\})^{\Z^2}$, with the two coordinates corresponding to vertical and horizontal channels
(while Red lines themselves do not cross each other,
this bit is also synchronised through other communication channels that \emph{do} cross other Red lines).
Notably, a Red corner necessarily encodes either $(\uparrow_H,\uparrow_V)$ or $(\downarrow_H,\downarrow_V)$.

In a (generically) all-Frozen tiling of the plane, we will encode one single infinite binary word.
Thus, we have a computable function $\gamma:\T_{\F_5}\to\{\uparrow,\downarrow\}^\N$
(defined on the generic all-Frozen set).
Now, we can remark that the pushforward operator 
$\gamma^*:\mathcal{M}_\sigma(\T_{\F_5})\to\mathcal{M}(\{\uparrow,\downarrow\}^\N)$
is actually an affine computable bijection.

\subsubsection{Controlling Words Through Entropy to Simulate Measures}
\label{sec:word-machine}

Now that we have the correspondence between 
the invariant measures on tiles
$\mathcal{M}_\sigma(\T_{\F_5})$ (in which $\G(\infty)$ will be included)
and the (non-invariant!) measures on words $\mathcal{M}(\{\uparrow,\downarrow\}^\N)$,
we need to implement a Turing machine
whose purpose will be to control
finite words on the binary alphabet $\{\uparrow,\downarrow\}$.

Each Blocking square in a marker
will contain an independent instance of the machine, with a space-time diagram of size $2^{3^k}$ that will perform its own computation. To do so, it will access the inputs:
\begin{itemize}
\item the scale $k$ of the Blocking square, already computed in Section~\ref{sec:blockable},
\item the first $b_\mathrm{read}(k)\simeq \log(k)$ bits of the word of length $3^k$ written in the hierarchy of Frozen Red squares
(seen as a Toeplitz encoding for the machine,
see \cite[Lemma 5.8]{GaySab23} for details on what it means and how to decode it),
\item a non-constrained binary seed $s$ of length $k$, that will introduce enough entropy to ``simulate'' the measures.
\end{itemize}
The encoded Turing machine $M$ will first check the length of $s$ to make sure $|s|=k$, then compute $M(k,s)\in\{\uparrow,\downarrow\}^{b_\mathrm{read}(k)}$, and finally check that $M(k,s)$ is indeed a prefix of the Toeplitz sequence
(if it isn't, then we will reach a forbidden state, an invalid tiling).
It follows that there are only $2^k$ different admissible computations,
one for each value of the seed $s$.
Notably, each of these computations must be done in less than $2^{3^k}$ steps to fit in the space-time diagram.

For such a Turing machine $M$,
we define $\T_{\F_6(M)}\subset \T_{\F_5}\times \A_6(M)^{\Z^2}$ the corresponding simulating SFT.

\subsection{Gibbs Measures for the Simulating Tileset} \label{sec:uniform-tileset}

From now on, the marker set $Q_k$ corresponds to the $\left(2\times 3^k+1\right)$-th scale of Robinson macro-tiles,
that are locally admissible with respect to $\F_6(M)$ on all the other layers.
We need appropriate sequences $(\epsilon_k)_{k\in\N}$ and $(\kappa_k)_{k\in\N}$
for Theorem~\ref{thm:equidistribution} to apply with overlapping intervals.

First, for the entropy criterion, we need to evaluate $\left|Q_k\right|$.
\begin{proposition}~\cite[Lemma 31, Proposition 42 and 43]{GaySabTaa23} \label{prop:cardinal}
We have $\abs{Q^{\symb{H}}_k}\simeq C_k^{16^{3^k}}$ with $2^{4^{-k}}\leq C_k\leq 2$, $\abs{Q^{\symb{B}}_k}\simeq \left(\abs{Q^{\symb{H}}_k}\right)^{\frac{3}{4}}$ and $\abs{Q^{\symb{B}}_k}\leq C^{4^{3^k}}$ for some $C>1$.
\end{proposition}
In particular, $\abs{Q_k}\approx\abs{Q^{\symb{H}}_k}$ and $\overline{\lambda_{Q_k}}\approx\overline{\lambda_{Q^{\symb{H}}_k}}$ as $k\to \infty$.

According to \cite[Lemma 45]{GaySabTaa23}, as a marker in $Q_{k+2}^{\symb{H}}$ is \emph{mostly} a grid of independent markers in $Q_k$, one has $\frac{\log(\abs{G_{n_{k+2}}})}{I_{l_{n_{k+2}}}}\geq
\frac{\log(\abs{Q_{k+2}})}{I_{l_{n_{k+2}}}} \geq
(1-\kappa_{k})\frac{\log(\abs{Q_k})}{l_k}$, with $\kappa_k=O\left(\frac{1}{t_k}\right)=O\left(\frac{1}{\log(k)}\right)$
corresponding to the density of non-Hot markers in the grid.

Now, taking $\epsilon_k=\sqrt{\frac{1}{l_k}}$, the temperature interval for the marker set 
$Q_k$ is
\[
T_k=\left[\frac{C}{\epsilon_k}\abs{I_{l_{n_{k+2}}}},C'\epsilon_k\frac{\abs{I_{l_k}}}{\abs{I_{l_k}}-\abs{I_{{l_k}-2}}}\right]\simeq\left[\frac{C\left(4^{3^k}\right)^2}{\epsilon_k},C'4^{3^{k+2}}\epsilon_k\right] \, .
\]
In particular, both bounds go to infinity,
and for sufficiently large scales $k$ the intervals $T_k$ and $T_{k+1}$ overlap
since $\frac{\max(T_k)}{\min(T_{k+1})}\underset{k\to\infty}{\longrightarrow}\infty$~\cite[Corollary 48]{GaySabTaa23}.

In other words, we satisfy the entropy and temperature criteria
for every marker set $Q_k$ in Theorem \ref{thm:equidistribution},
with temperature intervals that eventually overlap.
Now, after collapsing the two consequences of the theorem into 
\cite[Theorem 47]{GaySabTaa23},
we can conclude that for any $\mu\in\G(\beta)$ with $\beta\in T_k$, 
we have $\mu\approx \overline{\lambda_{Q_k^\symb{H}}}$,
with the weak-$*$ approximation getting increasingly tight as $k\to\infty$.

It follows that the model is uniform and $\G(\infty)=\acc{\overline{\lambda_{Q_k^\symb{H}}},k\to\infty}$,
so the remaining task is to understand what $\lambda_{Q_k^\symb{H}}$ looks like.

\subsection{Relating Uniform Markers to Measures on Words}
\label{sec:computable-words}

\newcommand{\Wl}{\mathcal{W}^l}

Remind that $\overline{\lambda_Q}$ is the invariant measure induced by a periodic grid of \emph{iid} markers uniformly distributed in $Q$.
We will here denote by $\Wl$ the event ``$0\triangleleft P^\symb{F}_{2l+1}$''
(\ie $0$ is included in a Frozen area encoding a word of length $l$ or more).
According Proposition~\ref{prop:frozen}, we have $\overline{\lambda_{Q^{\symb{H}}_{k}}}\left(\Wl\right)\underset{k\to\infty}{\longrightarrow}1$, so the conditional distribution
$\overline{\lambda_{Q^{\symb{H}}_{k}}}\left(\ \cdot\ \middle|\Wl\right)$ is an increasingly good approximation of $\overline{\lambda_{Q^{\symb{H}}_{k}}}$ for a fixed $l$.

For the uniform distribution
$\mathcal{U}\left(Q^\symb{B}_k\right)$,
the measure on words $w\in\{\uparrow,\downarrow\}^{b_\mathrm{read}(k)}$ induced by the central Blocking square is:
\[
m_k(w):=\frac{\abs{\{s\in\{0,1\}^{k}: M(k,s)=w \}}}{2^k} \, .
\]

A marker in $Q^{\symb{H}}_{k+1}$ is a grid
of markers in $Q_k$, with a proportion $\frac{1}{t_k}$ of $Q^{\symb{B}}_k$ and $\frac{t_k-1}{t_k}$ of $Q^{\symb{H}}_k$.
For the pushdown distributions by $\gamma_l$
(on words of length $l\leq b_\mathrm{read}(k)$),
noting $m_k^l$ the marginal of $m_k$ on the first $l$ bits, we have the following induction:
\begin{eqnarray*}
\gamma_l^*\overline{\lambda_{Q^{\symb{H}}_{k+1}}}\left(\ \cdot\ \middle|\Wl\right)&=&\frac{1}{t_k}\gamma_l^*\overline{\lambda_{Q^{\symb{B}}_k}}\left(\ \cdot\ \middle|\Wl\right)+\frac{t_k-1}{t_k}\gamma_l^*\overline{\lambda_{Q^{\symb{H}}_k}}\left(\ \cdot\ \middle|\Wl\right)\\
&=&\frac{1}{4t_k}m_k^l + \left(1-\frac{1}{4t_k}\right)\gamma_l^*\overline{\lambda_{Q^{\symb{H}}_k}}\left(\ \cdot\ \middle|\Wl\right)\\
&=&\frac{1}{4t_k}m_k^l+\left(1-\frac{1}{4t_k}\right)\left(\frac{1}{4t_{k-1}}m_{k-1}^l+\dots\right) \, .
\end{eqnarray*}

\begin{proposition}~\cite[Proposition 50]{GaySabTaa23}\label{proposition.ConditionalMeasure}
$\gamma_l^*\overline{\lambda_{Q^{\symb{H}}_{k+1}}}\left(\ \cdot\ \middle|\Wl\right)= \sum_{j=l}^k\frac{1}{4t_j}m_j^l\prod_{i=j+1}^k\left(1-\frac{1}{4t_i}\right)$.
\end{proposition}
Notably, if we repeat $m$ for all the $\left(m_j^l\right)_j\in\N$ after some scale,
we have $\gamma_l^*\overline{\lambda_{Q^{\symb{H}}_k}}\left(\ \cdot\ \middle|\Wl\right)
\underset{k\to\infty}{\longrightarrow} m$.

Now, remind that any $\Pi_2$-computable set $K\subset\M(\{\downarrow,\uparrow\}^\N)$
can be realised as an accumulation set for some computable sequence $\left(m_j\right)$.
As we are working with the weak-$*$ topology, without loss of generality,
we can restrict each $m_j$ to its first $j$ bits, to $\{\downarrow,\uparrow\}^j$.
Now, to indeed realise this $K$ as $\gamma^*\left(\G(\infty)\right)$,
we need to be slightly more careful, and work around two issues.
First, as pointed out in the previous paragraph, we need to repeat a given measure $m$
for many consecutive $m_j$ so that the actual measures 
$\gamma_l^*\overline{\lambda_{Q^{\symb{H}}_k}}$ can ``catch up'' to it.
This can be done explicitly by adding exponential repetitions,
\emph{i.e.} computing the sequence
$\left(m_{\left\lfloor \log_2(j)\right\rfloor}\right)_{j\in\N}$ instead.
Second, as underlined in Section~\ref{sec:word-machine},
the Turing machine can use ``only'' $2^{3^j}$
to generate the $j$-th measure of this sequence.
This can be worked around by trying to generate
the whole sequence $\left(m_j\right)$ at every scale $k$,
and halting the process once the allotted time $k$ is up,
which in practice amounts once again to adding some \emph{finite} number of repetitions
for each element of the sequence.
Further details of these computability arguments can be found
in~\cite[Section 7]{GaySabTaa23}.
With these tools in hand, we now have all the ideas
used in the proof of Theorem~\ref{thm:old}.

\newpage
\section{Proof of Theorem~\ref{thm:main}}

Now that we have explained the already existing construction for Theorem~\ref{thm:old},
let's delve into the proof of the new result, that expands this idea.

In the construction of Section~\ref{sec:old-paper},
to realise $\G(\infty)$ such that $\gamma^*\left(\G(\infty)\right)=X$
(with $X$ a $\Pi_2$-computable connected set),
we used a Turing machine $M_X$ that reads a binary seed $x$
of length $k$ (at the $k$-th scale of computations)
and outputs a binary word of length $b_\mathrm{read}(x)$.

Without loss of generality, we can assume we have access to \emph{several}
such binary inputs; we will thereon need a second one, $y$,
encoded as a string of length \emph{at most} $k$ followed by a blank symbol $\#$.
Quantitatively, going from a binary seed $x$ to a seed $(x,y)$ on a larger alphabet
marginally changes the exact quantitative values
in Proposition~\ref{prop:cardinal}
but without changing the overall qualitative exponential growth,
hence by the end of Section~\ref{sec:uniform-tileset},
the statement remains unaffected, we still have a uniform model,
with Gibbs measures in the neighbourhood of $\overline{\lambda_{Q_k^\symb{H}}}$
whenever $\beta\in T_k$.

As was the case in Section~\ref{sec:computable-words}, we will need to make sure all further computations involving $y$ stay bounded within the space-time diagram of size $2^{3^k}$.
As a rule of thumb, we will only consider operations in $\mathrm{PTIME}(k)$,
which avoids this issue for big-enough scales.

Notably, we will need to use a universal Turing machine $U$,
that can simulate any other Turing machine $M$ given an appropriate encoding $\left<M\right>$.
This simulation process can be done relatively efficiently,
so that simulating $M$ with $U$ only squares its initial complexity~\cite{NeaWoo06}.
In particular, if $M$ was in PTIME, so will be its simulation.
We now implement a ``universal'' Turing machine $U_X$ that \emph{first} reads $y$,
and if $y=\symb1^i\#^{n-i}$, then instead of simulating the machine $M_X$ on the input $x$,
it simulates $M_i$ \emph{the $i$-th Turing machine}
(for some fixed computable enumeration of all Turing machines).
Checking whether $|y|\leq k$ and if $y=1^i$ can be done with an automaton,
in at most $k$ steps, clearly in polynomial time.
For the simulation of $M_i$, we can always hardcode in $U_X$ a default behaviour
in case the computations take too long.

Without supplementary constraints, at a given scale $k$ of computations,
there is an increasingly low probability $\frac{k}{2^{k+1}-1}$
to \emph{not} simulate according to $M_X$.
It follows that the potential $\phi_X$,
associated to the embedding of $U_X$ in the simulating tileset
from Section~\ref{sec:simulating-tileset},
still induces $\gamma^*\left(G(\infty)\right)=X$ for its accumulation set.

Now, to force another accumulation set than $X$, it suffices to force $U_X$
to simulate a Turing machine \emph{other} than $M_X$.
More precisely, consider any other
connected $\pi_2$-computable set $Y$,
induced by a polynomial time Turing machine $M_Y$
in Theorem~\ref{thm:old}.
In particular, there is some $i\in\N$ such that $M_Y=M_i$,
so we need to force the value $y_i=\symb1^i\#^{\dots}$ for the input after some scale.
Notice that the ``sparse shape'' of the input tape embedded into a Robinson square is mostly the same from one scale to the next
(spread across an amount of actual tiles exponential in the length of the tape),
so we can in particular force $y=y_i$ with a local rule with forbidden patterns of size $r_i<\infty$.
We can thus define $\psi_Y$ the potential associated to these new local rules.

In particular, for the perturbed potential $\phi_X+\epsilon\psi_Y$ (with $\epsilon>0$),
Theorem~\ref{thm:equidistribution} still applies for all the corresponding marker sets $Q_k$,
which are the same regardless of the value of $\epsilon$, as the potential might use different weights depending on $\epsilon$ but always corresponds to the same forbidden patterns, the same tileset.
Notably, the range of the interactions $r_i$ and the intensity of the perturbations $\epsilon$
only affect the bounds of the temperature intervals $T_k$ by common multiplicative factors.
Following the rest of the proof for Theorem~\ref{thm:old},
we still have overlapping temperature intervals in \ref{sec:uniform-tileset}.
Notably, translating $y_i$ into $\left<M_i\right>$ in order to perform simulations might be
a very complex task computationally speaking,
but what matters here is that $i$ is \emph{fixed}, so this only adds a \emph{constant time}
overhead to the simulation of $M_i$ by $U_X$,
which is then still ultimately performed in $\mathrm{PTIME}(k)$.
In particular, we end up with $\gamma^*\left(\G(\infty)\right)=Y$ for the accumulation set,
which concludes the proof of Theorem~\ref{thm:main}.

\vfill

\sloppy
\hbadness=10000
\printbibliography

\end{document}